\documentclass[manuscript]{emulateapj}

\slugcomment{to appear in the Astrophysical Journal}

\shorttitle{Model of SMC3}
\shortauthors{Kato, Hachisu \& Miko{\l}ajewska}

\begin{document}


\title{AN X-RAY AND OPTICAL LIGHT CURVE MODEL OF THE ECLIPSING SYMBIOTIC 
BINARY SMC3
}


\author{Mariko Kato} 
\affil{Department of Astronomy, Keio University, Hiyoshi, Yokohama
  223-8521, Japan;}
\email{mariko@educ.cc.keio.ac.jp}


\author{Izumi Hachisu}
\affil{Department of Earth Science and Astronomy, College of Arts and
Sciences, University of Tokyo, Komaba, Meguro-ku, Tokyo 153-8902, Japan}

\and 

\author{Joanna Miko\l ajewska}
\affil{N. Copernicus Astronomical Center, Bartycka 18, 00-716 Warszawa, Poland}



\begin{abstract}
Some binary evolution scenarios to Type Ia supernovae include  
long-period binaries that evolve to symbiotic supersoft X-ray 
sources in their late stage of evolution. However, symbiotic stars with 
steady hydrogen burning on the white dwarf's (WD) surface are very rare,  
and the X-ray characteristics are not well known. 
SMC3 is one such rare example and a key object for understanding the evolution
of symbiotic stars to Type Ia supernovae.
SMC3 is an eclipsing symbiotic binary, consisting of a massive 
WD and red giant (RG), with an orbital 
period of 4.5 years in the Small Magellanic Cloud. 
The long-term $V$ light curve variations are reproduced 
as orbital variations in the irradiated RG, whose atmosphere 
fills its Roche lobe, thus supporting the idea that the RG supplies 
matter to the WD at rates high enough to maintain steady hydrogen burning on the WD.
We also present an eclipse model in which an X-ray emitting region 
around the WD is almost totally occulted by the RG swelling over 
the Roche lobe on the trailing side, although it is always 
partly obscured by a long spiral tail of neutral hydrogen 
surrounding the binary in the orbital plane.
\end{abstract}




\section{INTRODUCTION} \label{sec_introduction}

SMC 3 is a symbiotic star consisting of a massive white dwarf (WD) 
and an M giant.  Its orbital period is about 4.5 years.
It is a supersoft X-ray source (SSS) (RX J0048.4$-$7332) 
that has been under observation for 20 years \citep[][and references therein]
{jor96,mue97,kah04,kah06,ori07,stu11}. 
The WD appears to be very massive because  
SMC3 is the strongest X-ray source among symbiotic binaries 
\citep{mue97}.
In addition, a massive WD ($M_{\rm WD} > 1.18 M_\odot$) 
is suggested from the results of X-ray spectrum fittings with model atmospheres 
\citep{ori07}. 
The 20 years observation period is quite long compared with the  
durations of the supersoft X-ray phases of classical novae harboring massive WDs, 
0.7 years for 1.0 $M_\odot$, 2.7 years for 0.8 $M_\odot$, 
and 8.9 years for 0.6 $M_\odot$ \citep[see Table 2 in][]{kat94h}. 
Therefore, it is very unlikely that SMC3 is a nova remnant.  
\citet{ori07} concluded that SMC3 is a symbiotic binary with 
a steady hydrogen-shell-burning WD because no nova outbursts have been 
recorded in the last 50 years despite continuous optical monitoring 
of the SMC.

Steady hydrogen-shell-burning WDs have attracted attention in relation to 
Type Ia supernova (SN Ia) progenitors because these WDs accumulate accreted
matter, causing their masses to increase. 
In the single degenerate (SD) scenario to SNe Ia in which an accreting WD 
grows in mass to reach the Chandrasekhar mass and explodes as an SN Ia, 
there are two well-known paths of binary evolution  \citep{li97,hkn99,hknu99}: 
one is the main-sequence (MS) channel, for those with an MS companion,
and the other is the symbiotic channel, for those with an red giant (RG) companion. 
In these evolutionary channels, a typical binary evolves in its final stage  
from the wind phase (steady hydrogen-shell-burning with
optically thick winds) through the supersoft X-ray phase (steady 
hydrogen-shell-burning with no optically thick winds) to the recurrent nova phase 
\citep[intermittent hydrogen shell burning, e.g.,][]{hac10b}.
In the symbiotic channel, progenitor binaries are expected to 
evolve into symbiotic stars with a massive WD with steady
hydrogen-shell-burning \citep{hac12b}.  They are potentially 
strong SSSs. If they are detected, we use them as testbeds 
for the SN Ia progenitor scenarios. 
Unfortunately, symbiotic X-ray binaries with a massive WD  
are very rare. SMC3 is one such rare example.

It has been suggested that the supersoft X-ray fluxes of galaxies are much
weaker than expected for the SD scenario \citep{ori10,dis10a}.
\citet{dis10b} emphasized that the double degenerate (DD) 
scenario for SNe Ia, in which two WDs merge and then explode as an SN~Ia,
predicts a large number of symbiotic stars before the second
common envelope evolution that produces a DD system.
Therefore, the lack of supersoft X-ray flux is a problem common 
to both SD and DD scenarios.    
\citet{gil10} claimed that weak supersoft X-ray emission in early-type 
galaxies is negative evidence against the SD scenario. 
\citet{ori10} pointed out that most SSSs in M31 are 
transient, and steady-burning sources are rare, possibly because   
of low duty cycles of the SSS phases in these sources.  The work by \citet{gil10}
was also criticized by \citet{hac10b}, \citet{kat12iau}, and \citet{wan12},  
because Gilfanov \& Bogd\'an adopted not only overestimated numbers of expected SSSs
but also unrealistically large X-ray fluxes for symbiotic SSSs. 
If we use a reliable number of symbiotic SSSs based on the SD scenario 
\citep{hkn99} and realistic X-ray fluxes, we can statistically explain the weak 
supersoft X-ray emission in early-type galaxies, consequently, this  
strongly supports the SD model 
as a promising scenario for SNe Ia \citep{hac10b}.    
\citet{dil12} reported a detailed spectral analysis of 
the SN Ia PTF~11kx and concluded that it had a symbiotic nova progenitor, 
mainly because of a complex circumstellar environment.
\citet{chi12} also suggested that Kepler's SN (SN~1604) is Type Ia, and 
its progenitor was a symbiotic star consisting of a WD and an asymptotic
giant branch (AGB) star because of strong interaction with circumstellar
matter.

However, the nature of symbiotic steady-burning X-ray sources are yet 
to be fully understood; for example, their WD masses, RG companion masses, 
spectral types, orbital periods, wind mass-loss rates from both stars, and 
nebular emissions, as well as their origins and evolutionary paths
to the present binary state. 
In symbiotic stars, a hot WD may be embedded in a RG's cool wind 
that obscures supersoft X-rays \citep{hac10b,kat12iau,nie12}. 
For example, V407 Cyg, a symbiotic classical nova, showed a very weak 
supersoft X-ray flux in the later phase of its 2010 outburst because the WD 
is deeply embedded in a thick envelope originating from the cool 
giant wind \citep{sch11,nel12,hac12}.  Interestingly,  
strong symbiotic SSSs have been discovered so far only in the galactic halo (AG Dra) or 
in the Small Magellanic Cloud \citep[SMC3 and Lin358:][]{ori07}.  
This suggests that a low-metallicity environment is an important clue 
to the detection of supersoft X-rays from symbiotic systems.
To examine such a possibility, we need to study individual 
symbiotic SSSs in detail. 

In symbiotic binaries, a cool wind blows from the RG, and a hot wind blows from
the WD.  Radiation at various wavelengths may originate from 
different areas of the binary. Researchers have become interested mainly in 
SMC3's X-ray nature, but its binary nature has not yet been fully understood. 
In relation to binary evolution scenarios to SNe Ia, it is very important 
to know why the supersoft X-ray fluxes are so weak, 
although their WDs are believed to be very massive.    
In this work, we propose a binary model that explains both 
the long-term $V$ magnitude variations and observed X-ray eclipses.  
In Section \ref{sec_observation}, we summarize the observational 
properties of SMC3 and propose our basic model of it. 
In Section \ref{sec_Bmagmodel}, we present our model light curves 
for long-term variations in the $V$ magnitude. 
Section \ref{sec_model} presents our X-ray eclipse light curve model.  
Discussion and conclusions follow in 
Sections \ref{sec_discussion} and \ref{sec_conclusions}, respectively.

\section{THE MODEL OF SMC3 -- A SYMBIOTIC X-RAY SOURCE} \label{sec_observation}

\citet{mor92} reported low-resolution optical spectra of SMC3.  
The RG is classified as spectral type M0. 
The spectrum is typical for a high-excitation s-type 
symbiotic star and exhibits Balmer, \ion{He}{1}, and 
\ion{He}{2} emission lines as well as Raman-scattered 
\ion{O}{6} 6825 and 7082 features. 
The spectra also show an [\ion{Fe}{10}] emission line, 
an unusually high ionization line for a symbiotic star.
Optical spectra were also published by \citet{mue96} 
for 1994 Dec 20 -- 23 [orbital phase of $\phi=0.26$, 
see Equation (\ref{equation_ephemeris}) in Section \ref{sec_Bmagmodel}
below for the ephemeris], and by \citet{ori07} for 1994 Oct 17 ($\phi=0.22$). 
\citet{ori07} remarked that coronal lines 
[\ion{Fe}{9}] and  [\ion{Fe}{10}] in the optical spectra indicate 
a photoionizing source at a very high temperature of $T \sim 5 \times 10^5$ K.  
One {\it IUE} UV spectrum 
(SWP 47572L) was taken on 1993 April 30 (JD 2449107.5) \citep{vog94m} 
($\phi=0.89$, i.e., X-ray ingress).  
This spectrum appears typical for a symbiotic star of moderate 
excitation \citep{jor96}.  

Near-infrared (IR) $I$ magnitude light curves (shown later in Figure 
\ref{BVlightcurve.08}) taken by the OGLE II and III projects show an irregular
variation with a period about 110 days \citep{kah04,stu11},  
which is attributed to pulsation of the RG by \citet{kah04}. 
A blue filter light curve was secured by the MACHO project for as long as
6 years (also shown later in Figure \ref{BVlightcurve.08}); it shows
a long-term variation with a period of $\sim 1600$ days similar 
to the X-ray variation \citep{kah04,stu11}, superposed by a short-term
variation similar to that of the $I$ magnitude light curve.

\begin{figure}
\epsscale{1.15}
\plotone{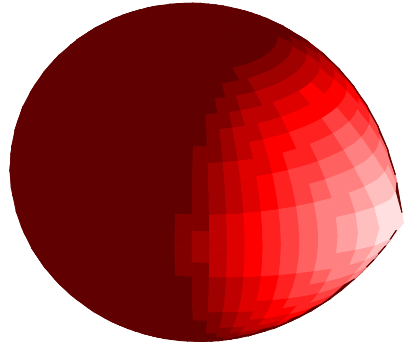}
\caption{Schematic view of the binary in the $V$ band. 
The RG fills its Roche lobe, and its hemisphere is irradiated 
by the hot WD (right: black dot). 
Distribution of the RG's surface temperature is indicated 
by red color gradient on a linear scale (white-red side is the highest 
temperature: 5400 -- 5500 K). 
The temperature of the RG on the non-irradiated side (dark red)    
is the parameter $T_{\rm RG}$ (3100 -- 3500 K). 
We assume a steady hydrogen-shell-burning WD. 
The WD and its surrounding hot gas are invisible in the $V$ band  
because of their high temperature. The RG tail of neutral hydrogen is 
unseen because it is transparent in the $V$ band. 
\label{binary.B}
}
\end{figure}

Therefore, as a basic model of SMC3,
we assume a binary consisting of a massive WD and an irradiated 
RG as plotted in Figure \ref{binary.B} (which is explained in detail 
in Section \ref{sec_RGradius}).
Table \ref{table_binary} gives the assumed masses of the WD and RG, the 
separation $a$, and the effective radii of the Roche lobes for
the RG and WD.  Here the binary separation $a$ is calculated from 
Kepler's third law, $a^3 =G(P_{\rm orb}/2 \pi )^2(M_{\rm WD} + M_{\rm RG})$, 
assuming a circular orbit and $P_{\rm orb}=1634$~days (as determined 
in Section \ref{sec_Bmagmodel}).  


\begin{deluxetable}{llllllll}
\tabletypesize{\scriptsize}
\tablecaption{Binary Parameters\tablenotemark{a}
\label{table_binary}}
\tablewidth{0pt}
\tablehead{
\colhead{WD mass} &
\colhead{RG mass} &
\colhead{} &
\colhead{ $a$} &
\colhead{$R_{\rm RL}$(RG)\tablenotemark{b}}&
\colhead{$R_{\rm RL}$(WD)\tablenotemark{c}}\\
\colhead{$(M_\odot$)} &
\colhead{$(M_\odot$)} &
\colhead{}&
\colhead{($R_\odot$)} &
\colhead{($R_\odot$)} &
\colhead{($R_\odot$)}
}
\startdata
1.35 & 3.0 & ...    & 953  & 428 & 298  \\
1.35 & 2.0 & ...    & 873  & 361 & 302  \\
1.35 & 1.5 & ...    & 828  & 321 & 306  \\
1.35 & 1.0 & ...    & 776  & 274 & 314  \\
1.35 & 0.8 & ...    & 753  & 252 & 320  \\
1.35 & 0.6 & ...    & 729  & 227 & 329  \\
1.35 & 0.4 & ...    & 703  & 197 & 342  \\
1.20 & 0.8 & ...    & 735  & 253 & 305  \\
1.10 & 0.8 & ...    & 723  & 254 & 294  \\ 
1.00 & 0.8 & ...    & 710  & 255 & 283  

\enddata
\tablenotetext{a}{We assume $P_{\rm orb} =1634$ days and a circular orbit.}
\tablenotetext{b}{Effective Roche lobe radius of the RG \citep{egg83}}
\tablenotetext{c}{Effective Roche lobe radius of the WD \citep{egg83}}
\end{deluxetable}

\subsection{The RG Component}\label{sec_RGradius}

Near-IR photometry of SMC3 is reported as 
$\langle J\rangle =11.935$, $\langle H\rangle =11.035$, and 
$\langle K_{\rm s}\rangle =10.8$ at JD 2451034.7 \citep{phi07}
(hereafter $K_{\rm s}=K$). 
This date is close to mid-eclipse, $\phi=0.07$.   
These $JHK$ magnitudes represent the RG magnitudes well. 
Then, we obtain reddening-corrected magnitudes of 
$J_0=11.848$, $H_0=10.977$, and $K_0=10.765$ with  
$A_J=0.874 \times E(B-V)=0.087$, 
$A_H=0.589 \times E(B-V)=0.058$, 
and $A_K=0.353 \times E(B-V)=0.353 \times 0.099 = 0.035$.
Here we use the extinction coefficients given by \citet{car89} and the 
extinction $E(B-V)=0.099$ given by \citet{hil05}. 
In our light curve fittings, on the other hand, the RG magnitudes are
obtained from the light curve minimum; i.e., when the RG is in front of the WD,
we see the non-irradiated side of the RG. 
From Figure \ref{BVlightcurve.08} we get $I \sim 13.34$ and $V \sim 15.35$.
Thus, the reddening-corrected magnitude becomes $I_0=13.193$  
and $V_0=15.043$, with $A_I=1.48 \times 0.099= 0.147$ and
$A_V=3.1 \times 0.099=0.307$ \citep{car89}. 

The RG luminosity is estimated using the relationship  
between $BC(K)$ and $(J-K)_0$ obtained 
for SMC variables \citep{bes84},  
\begin{eqnarray}
BC(K)&=&0.60+2.65 (J-K)_0 - 0.67 (J-K)_0^2, \cr
& &~~~{\rm for~} 0.6 < (J-K)_0 < 1.5.
\label{equation_bck}
\end{eqnarray}
We obtain $BC(K)=2.68$ for $(J-K)_0 =1.083$.  
The bolometric magnitude becomes $M_{\rm bol}=10.765+2.68-18.91=-5.46$,  
i.e., $L_{\rm bol}= 12000~ L_\odot$. Here we use  
$M_{\rm bol}({\rm Sun})=4.75$ for the absolute bolometric luminosity of the Sun.
\citet{bes84} also presented another relationship, $BC(K)$ vs. $(V-K)_0$, from which 
we get $BC(K) \sim 2.7$; this is quite consistent with the 
estimate obtained from Equation (\ref{equation_bck}). 

The effective temperature of the RG is estimated using the color-temperature 
relation of a model atmosphere for low-metallicity RGs given by
\citet{ku06} together with a metal abundance of [M/H]$=-1$.  
We get $T_{\rm eff} \sim$ 3800 K for $(J-K)_0=1.083$, 
$T_{\rm eff} \sim$ 3600 K for $(V-K)_0=4.278$, and 
$T_{\rm eff} \sim$ 3700 K for $(V-I)_0=1.85$. 
These color temperatures are consistent with each other within the 
accuracy of the model temperature ($\pm 100$ K). 
They are also consistent with the spectral type M0 estimated
by \citet{mor92} because for Galactic bright giants, the spectral type M0
corresponds to 3700 - 3800 K \citep{str81}. 
The radius of the RG becomes $250~R_\odot$ for $T_{\rm eff}=3800$ K, 
$270~R_\odot$ for 3700 K, and $280~R_\odot$ for 3600 K, 
with Stefan-Boltzmann's law $L=4 \pi R_{\rm RG}^2 \sigma T_{\rm eff}^4$. 
Comparing these values with the Roche lobe radii in Table 
\ref{table_binary} (i.e., 252 $R_\odot$ for a $0.8~M_\odot$ RG and 
274 $R_\odot$ for a $1.0~M_\odot$ RG), we see that 
the estimated RG radii are quite consistent with the effective Roche lobe radii  
of the RG.  Thus, we may safely assume a Roche-lobe-filling RG (Figure \ref{binary.B}). 

The RG is a semiregular variable with a pulsation period of 110 days. 
SMC3 is located in the upper part of the period - $K$ magnitude 
diagram for SMC OGLE variables  
\citep{ita04}. In this diagram, SMC3 is located at the border between 
sequences A$^+$ and B$^+$, both of which are classified as ``less regular 
pulsation AGB variables.''  
The pulsation mode is probably not fundamental, but is likely the first/second 
or third overtone. This is consistent with the irregular $I$ band variation 
with a small amplitude of $\Delta I \sim 0.05$ mag.


\begin{figure}
\epsscale{1.15}
\plotone{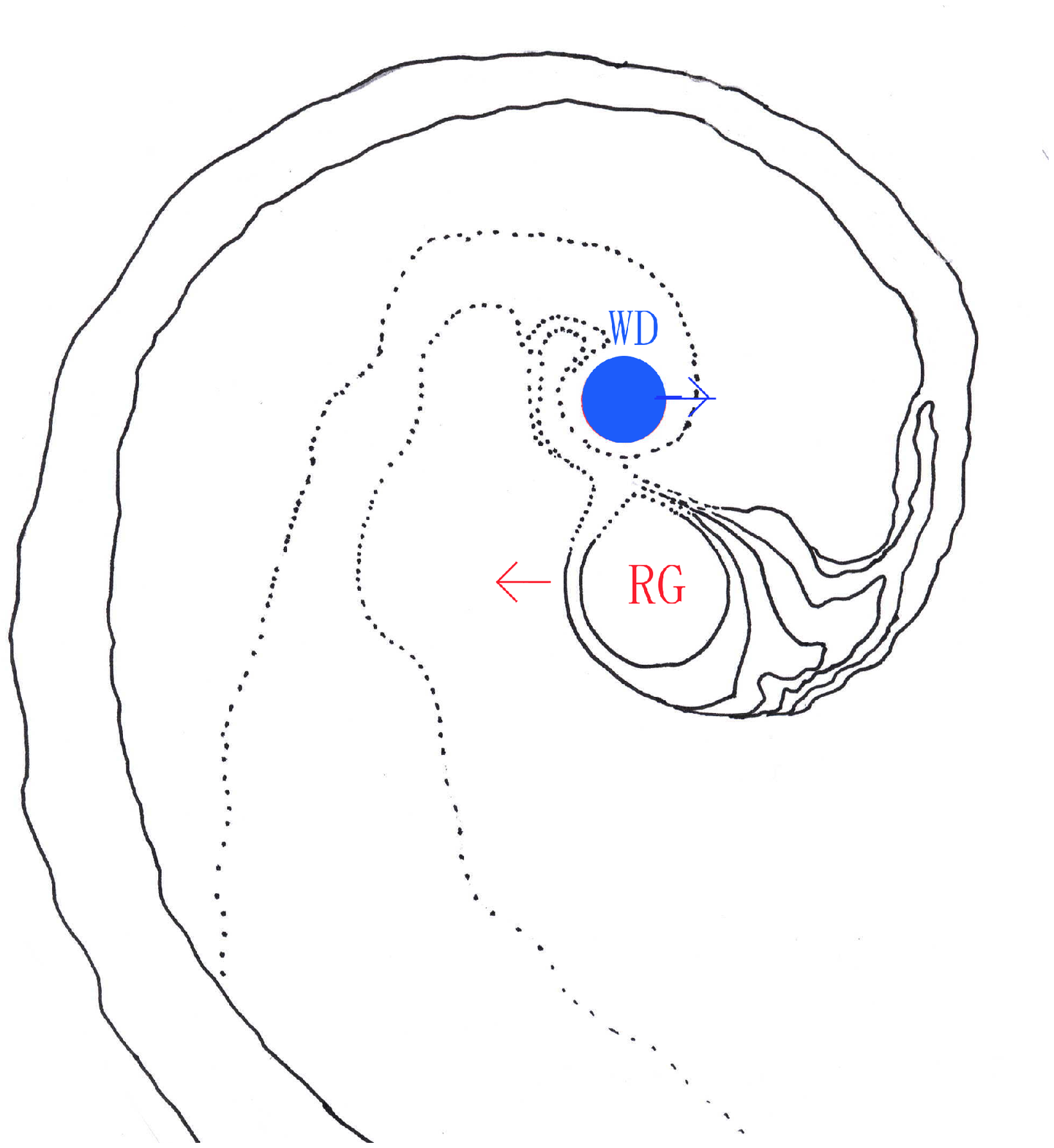}
\caption{ Schematic polar view of SMC3. 
Blue circular region represents a hot ionized plasma centered 
on the WD that scatters supersoft X-ray photons from the WD. 
The RG almost fills its Roche lobe and has a large trailing tail 
of neutral hydrogen that absorbs supersoft X-rays. 
Arrows denote the orbital motion of each component. 
Ionized and neutral regions are plotted by dotted and solid lines, respectively. 
This schematic figure was made using information from numerical
calculations \citep{saw86,moh12}. Contours denote the density taken from 
\citet{saw86}. Size and shape of each part are not important in our 
X-ray light curve model except for the height (thickness) from the orbital plain.  
\label{binary.X}
}
\end{figure}


\begin{deluxetable}{llll}
\tabletypesize{\scriptsize}
\tablecaption{Light Curve Model of X-Ray Eclipse 
\label{table_eclipse}}
\tablewidth{0pt}
\tablehead{
\colhead{Subject} &
\colhead{} &
\colhead{Model 1} &
\colhead{Model 2} 
}
\startdata
Duration of total X-ray eclipse $\Delta \phi$& ... &0.22  & 0.19  \\
Duration of total X-ray eclipse (day)& ... & 363 ~day & 304~day \\
Delay of X-ray mid-eclipse $\Delta \phi$& ... &0.075  & 0.075  \\
Delay of X-ray mid-eclipse (day) & ... &123~day  & 123~day   \\
RG radius             & ...& 245 $R_\odot$    &199 $R_\odot$  \\ 
Minimum height of the tail & ...&48~$R_\odot$ & 31~$R_\odot$ \\
Radius of the hot nebula $R_{\rm NE}$ & ...&100~$R_\odot$&100~$R_\odot$ 
\enddata
\end{deluxetable}

\subsection{The Hot Component: Supersoft X-ray Source} 

As briefly introduced in Section \ref{sec_introduction},  
SMC3 (=RX J0048.4$-7332$) is an exceptionally bright SSS 
among symbiotic stars \citep[see Table 1 in][]{mue97}, and 
has been observed with {\it ROSAT}, {\it Chandra}, and {\it XMM-Newton} 
\citep[][and references therein]{jor96,kah04,ori07,stu11}. 
\citet{jor96} analyzed {\it ROSAT} spectra combined with a UV spectrum
from {\it IUE}.  They found that a very hot Wolf-Rayet-type model atmosphere
fit the spectra reasonably well, but blackbody and hydrostatic
plane-parallel models did not.  In their analysis, the wind mass-loss rate
is several times $10^{-6}M_\odot$~yr$^{-1}$ and the central stellar surface
is hidden by stellar winds.  
However, they combined the {\it ROSAT} ($\phi=0.65$: out of eclipse) 
and the {\it IUE} spectrum ($\phi=0.89$: ingress), which are not  
simultaneous, so they may have overestimated 
the wind mass-loss rate in order to reproduce the smaller UV flux 
(due to the half-eclipsed state) with only the wind absorption.


The supersoft X-ray count rates for SMC3 decreased by a factor of 50 over 
more than 300 days starting in May 1993 \citep{kah04}. Kahabka 
presented a model of this X-ray decrease in which 
the WD orbits within a massive wind from the RG. The wind is highly ionized  
but still contains neutral hydrogen. The column density of neutral
hydrogen ($N_{\rm H}$) 
in the line of sight changes periodically with the orbital motion of the WD. 
As a result, the X-ray count rate decreases when the WD is on far side 
as seen from the Earth. 
\citet{ori07} criticized this scenario because this model predicts  
nebular emission lines at the high mass-loss rate of the RG wind they assumed, 
but no such lines were observed in the optical spectra. 
Orio et al. argued that the values of $N_{\rm H}$ appeared to be 
unchanged or only slightly higher in the 2003 eclipse, 
which seemed to be consistent with a real eclipse caused by the 
RG itself, rather than with a significant contribution 
by additional obscuration caused by an RG wind.

Recently, \citet{stu11} analyze spectra newly obtained 
with {\it XMM-Newton} EPIC-pn, using a blackbody model, 
in both high and low X-ray flux stages. 
Their spectral fits do not yield good results for a model of 
changing neutral-hydrogen column density 
because the spectra in low X-ray flux stages cannot be reproduced. 
Therefore, Kahabka's model was not supported. 
Instead, they suggested Compton scattering by fully ionized matter
that is distributed azimuthally unevenly in the orbital plane.
This requires the RG cool wind to completely ionize up to the RG
surface because the neutral-hydrogen column-density does not increase. 
However, such a situation is very unlikely because Raman-scattered light
was observed \citep{mor92}.
\citet{stu11} also suggested a temperature variation model for the eclipse, but 
this seems to be inconsistent with the light curves. If the temperature varies 
while the total luminosity remains constant, the  
$V$ magnitude light curve should be anti-correlated  
with the X-ray variation, but 
there is no indication of such behavior. Although these authors misunderstood and
rejected, the rest of their models, i.e.,  the varying emitting area model
with constant temperature and column density,
corresponds to the ``real eclipse,'' as \citet{ori07} already suggested, 
and reproduces very well the spectra at both high and low X-ray fluxes.

On the bases of these analyses, we assume that
the WD is embedded in very hot, highly ionized gas 
that scatters supersoft X-ray photons emitted from the WD. 
In other words, the X-ray emitting region is not a point source  
but is substantially extended around the WD, as illustrated by the blue
circular region in Figure \ref{binary.X}. 
We also add circumbinary matter distributed as in Figure \ref{binary.X},  
in which the RG swelling over its Roche lobe 
has a long spiral tail of neutral hydrogen
(see Section \ref{sec_model} for details).

\section{LIGHT CURVES OF $V$ AND $I$ VARIATIONS}\label{sec_Bmagmodel}

We assume a binary configuration as in Figure \ref{binary.B}  
to calculate the $V$ and $I$ light curves.  
As we already discussed in Section \ref{sec_RGradius}, 
it is very likely that the RG almost fills its Roche lobe. 
We assume a circular orbit with an inclination angle of $i=90 ^\circ$ and WD
and RG masses of 
$M_{\rm WD} =$ 1.0--$1.35~M_\odot$ and $M_{\rm RG} =0.8$-$2.0~M_\odot$, respectively. 
We further assume the RG radius to be constant, although the RG 
is pulsating as a semiregular variable. 
This approximation is good if the amplitude of the radius pulsation is small.  
In the symbiotic nova PU Vul, the companion RG is also a semiregular variable,  
and the amplitude of its radius pulsation is estimated to be 3--7\% 
by eclipse analysis \citep{kat12}. Because the amplitude of $V$ variation 
in SMC3 ($\sim 0.15$ mag) is much smaller than that in PU Vul ($\sim 1$ mag), 
a constant RG radius is a good approximation.

The WD has hydrogen shell burning so its surface 
is hot enough to emit supersoft X-rays.  The hemisphere of the RG
toward the WD absorbs high-energy photons emitted from the WD, and
this side is heated (see Figure \ref{binary.B}).  The heated surface emits
blackbody radiation at a local temperature calculated from the energy
balance. In contrast, in the opposite hemisphere,
the temperature remains low at the original surface temperature,
$T_{\rm RG}$, which we take as a parameter.
We neglected the temperature change due to RG pulsation.  
The RG surface is divided into small patches ($N=N_r \times N_\theta = 
64 \times 32$) and the energy balance is calculated at each patch to 
obtain its local temperature. 
Then, we sum them up to obtain the total $V$ and $I$ magnitudes of the RG 
at a specified orbital phase. 
A detailed description of our calculation method is given in \citet{hac01}. 

The WD and its hot thin nebula (blue region in Figure \ref{binary.X}) 
are extremely hot \citep[$T \sim 5 \times 10^5$ K;][]{ori07}, 
so their emission contributes little to the $V$ and $I$ bands. 
We considered the hot nebula surrounding the WD to be optically thin 
(see the discussion in Section \ref{sec_discussion}),  
so it is transparent in the $I$ and $V$ bands.   
We also assumed that the spiral tail is neutral and transparent 
in these $I$ and $V$ bands, as described in Section \ref{sec_config}. 

However, the cool tail may contribute to the $I$ magnitude.
As the standard model in this 
work, we determined the original RG temperature (i.e., that of the non-irradiated surface)
by fitting our $V$ model light curves with the observed $V$ magnitude. 
The resulting $I$ light curve is darker than the observed one and does not fit 
well, as seen below. 
Therefore, we added a nebular contribution in the $I$ band in order to fit 
our model light curve with the observed $I$ band light curve. 
We suppose that this nebular contribution is due mainly 
to the cool RG tail.

We assumed a steady hydrogen-shell-burning WD of bolometric luminosity 
$L_{\rm WD}= 3.77 \times 10^4~ L_\odot$ for $M_{\rm WD}=1.35~M_\odot$,
$L_{\rm WD}= 3.59\times10^4~L_\sun$ for $M_{\rm WD}=1.2~M_\odot$, 
$L_{\rm WD}= 3.34\times10^4~L_\sun$ for $M_{\rm WD}=1.1~M_\odot$, 
and $L_{\rm WD}= 3.02\times10^4~L_\sun$ for $M_{\rm WD}=1.0~M_\odot$.
We also assumed the RG mass to be $M_{\rm RG}=0.8,~1.0,~1.5$ 
and $2.0~M_\odot$.  
SMC3 is located in the direction of a dense \ion{H}{1} cloud of 
$N_{\rm H}=4 \times 10^{21}$ cm$^{-2}$ \citep{kah96}, but 
SMC3 is in front of this cloud, because the column density  
derived from spectral analysis of supersoft X-rays is much smaller 
than this value \citep[$N_{\rm H} \sim $  several $\times 10^{20}$ cm$^{-2}$:]
[]{ori07,stu11}.  Therefore, we adopted $E(B-V)=0.099$,
a distance modulus of 18.91 (a distance of 60.6 kpc), $A_{\rm I}=0.179$,
and $A_{\rm V}=3.1 \times E(B-V)=0.31$ after \citet{hil05}.


\begin{figure}
\epsscale{1.15}
\plotone{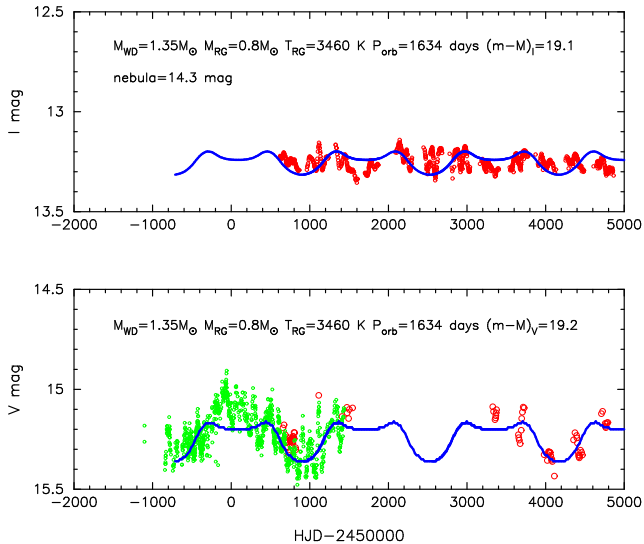}
\caption{$I$ and $V$ light curves of SMC~3.  Solid curves  
denote our model light curves for a 1.35 $M_\odot$ WD and 
0.8 $M_\odot$ RG. The original temperature of 
the RG (i.e., the temperature on the non-irradiated surface), $T_{\rm RG}$, 
is the only parameter determined from light curve fitting in $V$ band.  
We add a nebular emission of $m_I=14.3$ to fit the $I$ band data.
The $I$ and $V$ data were taken from 
OGLE II/III (red open circles) and MACHO (small green open circles).  
Our light curve model reproduces well the long-term variation in the $V$ magnitude 
as well as the very small dynamic range ($\sim 0.1$ mag)
of $I$ magnitude variation. 
Irregular deviation from our model, especially in the $I$ magnitude, may 
originate from the RG pulsation, possibly in a higher overtone mode, which 
our model does not include.  
\label{BVlightcurve.08}
}
\end{figure}

Figure \ref{BVlightcurve.08} shows the OGLE-II \citep{uda97} and
OGLE-III\footnote{http://ogle.astrouw.edu.pl/}  
\citep{uda08} $I$ and $V$ bands as well as the 
MACHO\footnote{http://wwwmacho.unu.edu.au/} 
blue band light curve. 
The MACHO light curve was shifted in magnitude to match OGLE's $V$ 
magnitude.  This figure shows long-term variations in the $I$ and $V$ magnitudes,
both of which are superposed with short pulsational variations of the RG. 

Our model light curves are shown in Figure \ref{BVlightcurve.08} for 
a 1.35 $M_\odot$ WD and a 0.8 $M_\odot$ RG. 
When the RG is in front of the WD ($\phi \sim 0.0$), we see only 
the original surface of the RG, where the magnitudes are faintest.
On the other hand, when the RG is on the far side of the orbit 
($\phi \sim$ 0.3 - 0.7), 
we see the irradiated bright hemisphere of the RG. Thus, the light 
curve shows a wide dip. Secondary minima exist, as clearly shown 
in our model $I$ light curve.  
They are due to ellipsoidal variation (a double peak in one orbital period) 
of a lobe-filling companion.
Here we assume that  only the RG contributes to the $V$ band emission, 
not the nebulae, the contribution of which is discussed later.  
We found that the RG surface temperature of the non-irradiated side 
$T_{\rm RG}= 3460$ K yields a good result in fitting with the $V$ light curve. 
To fit the $I$ band light curve,
we had to assume a nebular contribution of $m_{\rm I}=14.3$.

If we assume no nebular contribution to $I$, we need to assume a higher 
temperature, $T_{\rm RG}= 3700$ K, to obtain a good result.
This may be explained if we are observing different ``photospheres'' of the 
RG envelope at different wavelengths. When we observe the $I$ magnitude, 
we may see a deeper region where the temperature is higher. 
Another possible explanation is contamination by emission 
from the cool tail. The tail may contribute to the $I$ band, but not 
the $V$ band, 
reducing the temperature difference between different bands. 
Although the $I$ band is contaminated by semiregular pulsations, 
our $I$ light curve adequately reproduces the very small dynamic range
($\sim 0.1$ mag) of the observed light curve. 


\begin{figure}
\epsscale{1.15}
\plotone{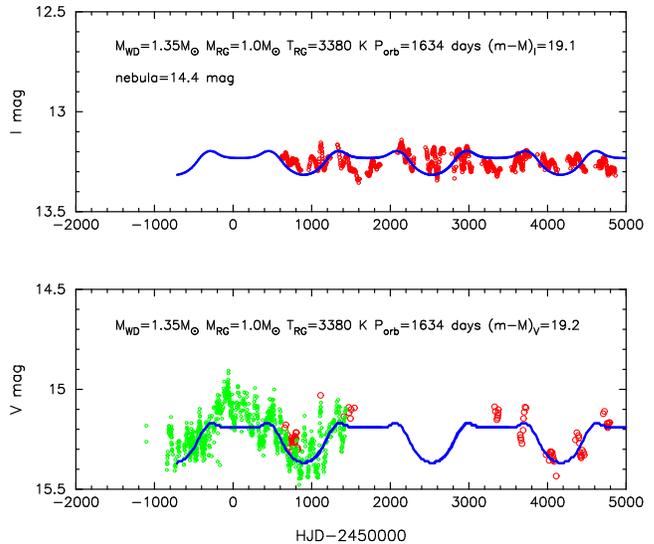}
\caption{Same as Figure \ref{BVlightcurve.08}, but for 
a 1.0 $M_\odot$ RG. 
We added nebular emission of $m_I=14.4$ to fit the $I$ band data.
\label{BVlightcurve.10}
}
\end{figure}


\begin{figure}
\epsscale{1.15}
\plotone{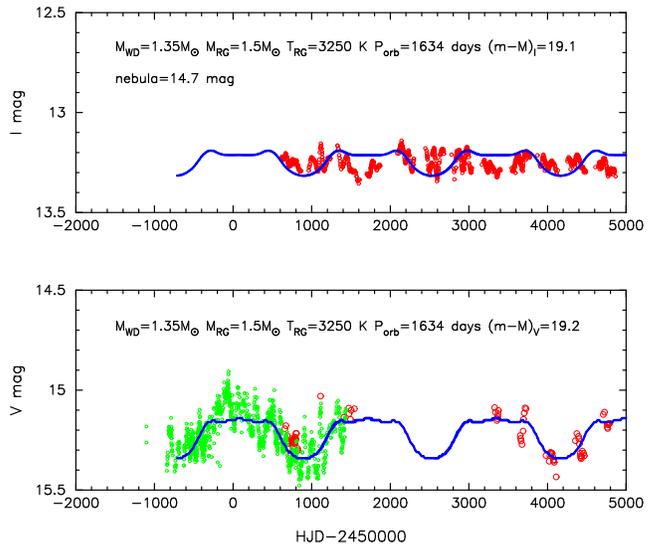}
\caption{Same as Figure \ref{BVlightcurve.08}, but for 
a 1.5 $M_\odot$ RG. 
We added nebular emission of $m_I=14.7$ to fit the $I$ band data.
\label{BVlightcurve.15}
}
\end{figure}


\begin{figure}
\epsscale{1.15}
\plotone{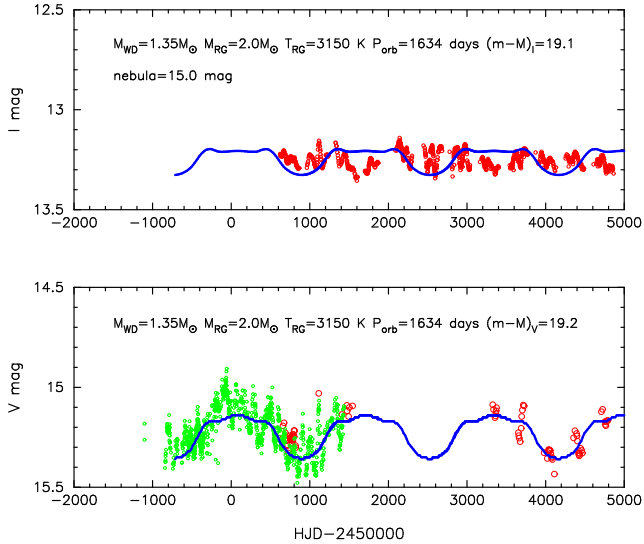}
\caption{Same as Figure \ref{BVlightcurve.08}, but for 
a 2.0 $M_\odot$ RG. 
We added nebular emission of $m_I=15.0$ to fit the $I$ band data.
\label{BVlightcurve.20}
}
\end{figure}

Figures \ref{BVlightcurve.10}--\ref{BVlightcurve.20} show  
similar model light curves for a 1.35 $M_\odot$ WD with 
different RG masses, i.e., $M_{\rm RG}= 1.0,~1.5$, and $2.0~M_\odot$. 
For a higher RG mass, we need to assume a lower RG surface temperature 
in order to obtain a similar magnitude.  For a more massive RG, 
we obtained a larger separation $a$, and thus a larger effective Roche lobe
radius (see Table \ref{table_binary}).  Therefore, we had to assume
a lower temperature in order to obtain a similar magnitude. 
For each model, we assume a nebular contribution of $m_{\rm I}$=14.3--15 mag. 

As in the previous section, the RG pulsation is not a fundamental mode, 
but possibly exists in higher overtones, which is consistent with the 
irregular pulsation behavior at small amplitude ($\sim 0.05$ mag).
These pulsations are not included in our theory, so we cannot reproduce
the complicated $I$ band variations. Instead, it is important 
that our light curve model reproduces 
the upper and lower limits of the $I$ magnitude around the mean value 
by the same model that reproduces the long-term variation in the $V$ band.  
In this sense, we may conclude that our model is consistent with both the 
$V$ and $I$ observations.

In Figures \ref{BVlightcurve.08}--\ref{BVlightcurve.20}, 
the $V$ magnitude is reproduced well by all of our 
theoretical models, whereas the
$I$ magnitude may be better reproduced by a low-mass RG companion. 
As shown in the previous subsection, 
$T_{\rm RG}= 3700 \pm 100$ K is quite consistent with the  
pulsation theory, so we prefer a 0.8 M$_\odot$ RG model to the others.  

\begin{figure}
\epsscale{1.15}
\plotone{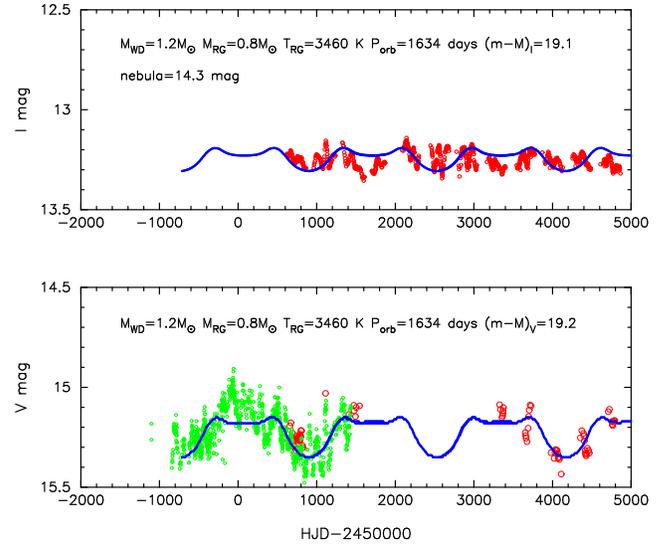}
\caption{Same as Figure \ref{BVlightcurve.08}, but for 
a 1.2 $M_\odot$ WD.
\label{BVlightcurve.wd120}
}
\end{figure}

\begin{figure}
\epsscale{1.15}
\plotone{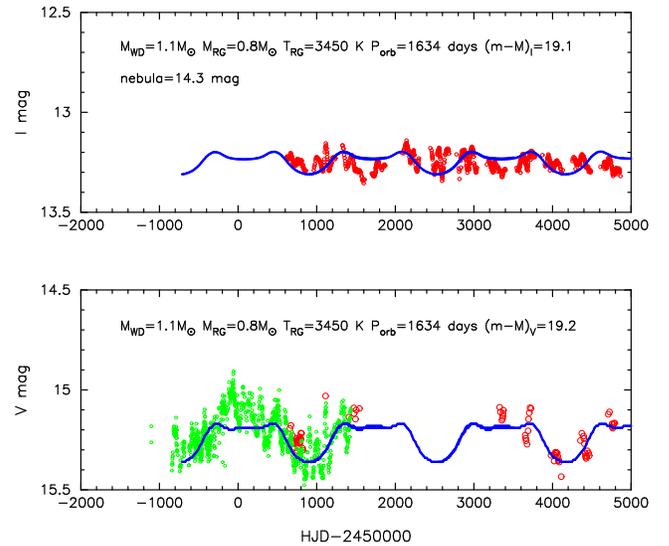}
\caption{Same as Figure \ref{BVlightcurve.08}, but for 
a 1.1 $M_\odot$ WD.
\label{BVlightcurve.wd110}
}
\end{figure}

\begin{figure}
\epsscale{1.15}
\plotone{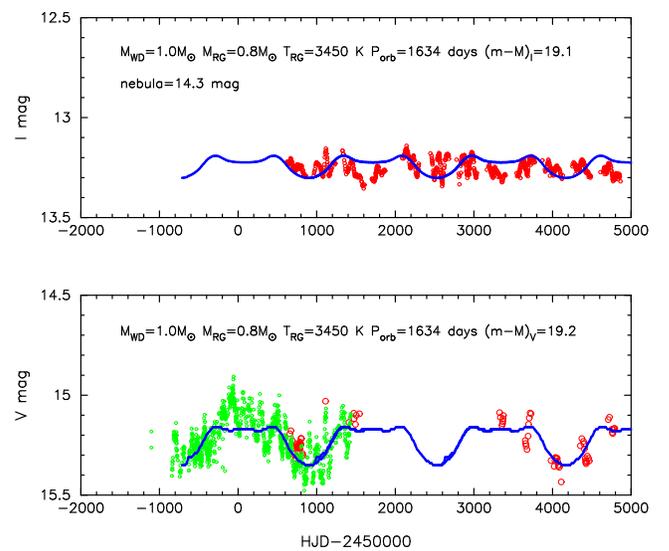}
\caption{Same as Figure \ref{BVlightcurve.08}, but for 
a 1.0 $M_\odot$ WD.
\label{BVlightcurve.wd100}
}
\end{figure}

We also calculated $V$ and $I$ light curves for different masses 
of the WD, i.e., $M_{\rm WD}=1.2$, 1.1, and 1.0 $M_\odot$, which are
more or less similar in both the $V$ and $I$ magnitudes 
(Figures \ref{BVlightcurve.wd120}, \ref{BVlightcurve.wd110},
and \ref{BVlightcurve.wd100}, respectively). 
Because both the $I$ and $V$ light curves are contaminated
by irregular RG pulsations, we cannot determine which is the best
fit to the observation.  Thus, we could not determine the WD mass
(or constrain the range of the WD mass) using only these light curve fittings.


\begin{figure}
\epsscale{1.15}
\plotone{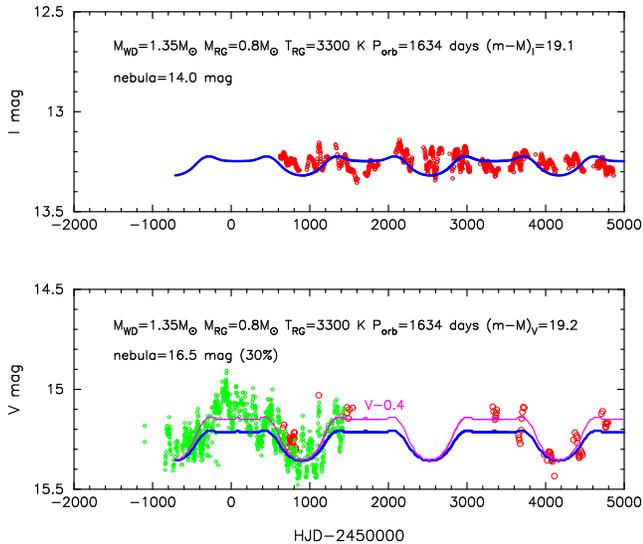}
\caption{Same as Figure \ref{BVlightcurve.08}, but with 
nebular emission of $m_V=16.5$ and $m_I=14.0$ (thick solid blue lines). 
Thin solid magenta line denotes the original $V$ magnitude (without nebulosity)
shifted upward by $-0.4$ mag.  The nebular emission of $m_V=16.5$
corresponds to about 30\% of the original luminosity.  The calculated
eclipses become too shallow to reproduce the observation.
\label{BVlightcurve.wd135nebula}
}
\end{figure}

The outer part of the hot thin nebula around the WD could 
contribute to the $V$ band because it has a large volume even though
it has very low emissivity.  Part of the cool RG tail could also 
contribute to the $V$ band if it is ionized by irradiation.
Thus, we calculated an additional light curve model in which 
the nebular contribution to the $V$ band is as much as  
30\% of that of the RG. 
Figure \ref{BVlightcurve.wd135nebula} shows such a case for 
a pair of 1.35 $M_\odot$ WD and 0.8 $M_\odot$ RG. 
For example, if the nebular emission amounts to 30\% of the RG luminosity in the $V$ band, 
we obtain a lower temperature of $T_{\rm RG}=3300$~K for the 
RG companion.  The total $V$ magnitude is shown by a blue line.
To fit our model with the $I$ magnitude at the same RG temperature,
we must increase the nebular contribution from $m_{\rm I}=14.3$
to $m_{\rm I}=14.0$.
If we compare this model with Figure \ref{BVlightcurve.08}, the amplitude
between minima and maxima becomes smaller in the $V$ band, which is 
inconsistent with the observations.
Therefore, we may say that the nebular contribution 
is probably less than $m_{\rm V}~> ~16.5$. 

In many symbiotic stars, the $I$ and $V$ band magnitudes are strongly 
contaminated by nebular emission \citep{nus87,sko98,sko05,sko09}. 
In SMC3, however, the companion RG is a very bright AGB star. 
This explains why the nebular contributions in $I$ and $V$ 
are rather small.

In our $V$ light curve fitting, we use the ephemeris
\begin{equation}
{\rm MJD}_{\rm min,V} = 49280 + N \times (1634)~ {\rm days}.
\label{equation_ephemeris}
\end{equation}
This is consistent with those of 
JD$_{\rm min}= 2449360 + N \times (1600\pm 140) $ days by \citet{kah04}, and 
MJD$_{\rm min,B}= (49242 \pm 9) + N \times (1647 \pm 24)$ days by  
\citet{stu11}.

Periodic variations similar to those of SMC3 were observed in YY Her, 
a symbiotic binary of $P_{\rm orb}=590$ days. 
The RG fills its Roche lobe, and its hemisphere toward the 
WD is heated by irradiation. 
\citet{mik02} showed that the periodic change in the $UBVRI$ magnitudes 
can be described by a combination of ellipsoidal and sinusoidal changes.
The RG radiation dominates in longer wavelength bands
among $UBVRI$, so 
the variation can be represented well by ellipsoidal changes. 
On the other hand, the nebular continuum dominates in shorter wavelength bands,   
so the light curve shows a single minimum in one orbital period.  
These characteristic properties are already shown in 
our model light curves for SMC3. 
Figures \ref{BVlightcurve.08}--\ref{BVlightcurve.20}
show that the $I$ magnitude variations have double peaks 
whereas the $V$ light curves display  
a single eclipse because the secondary minimum is compensated by 
extra radiation from the irradiated side of the RG. 


\begin{figure}
\epsscale{1.15}
\plotone{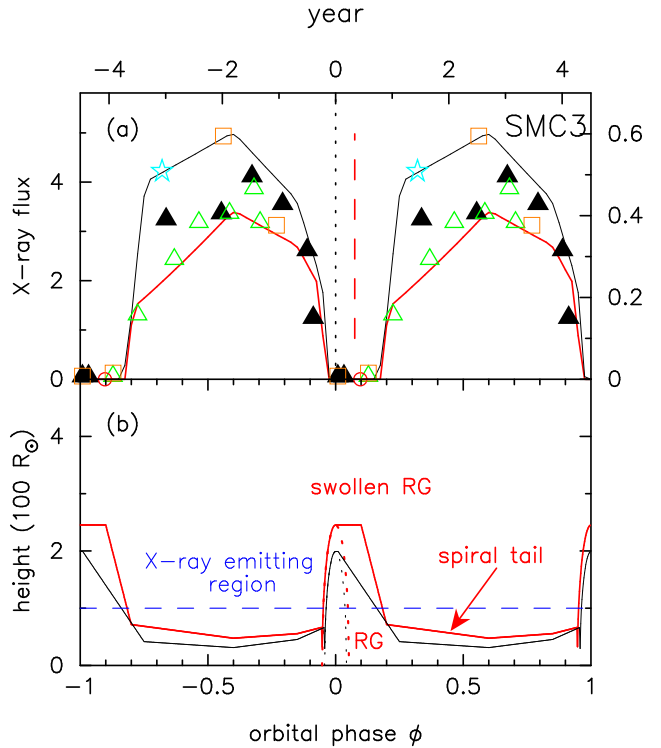}
\caption{
(a) X-ray light curve of SMC3. Data are taken from \citet{stu11}.   
Filled triangles: {\it ROSAT} PSPC. Open triangles:
{\it ROSAT} HRI. Open circles: {\it Chandra}.  
Open squares: {\it XMM-Newton}. Stars: {\it Swift}.
The supersoft X-ray flux (0.2 -- 1.0 keV) 
is in units of $10^{-12}~$erg~cm$^{-2}~$s$^{-1}$ (left ordinate). 
Solid lines indicate model light curves of X-ray eclipses;
Model 1 for $R_{\rm RG}=245~R_\odot$ (thick red line), Model 2 
for $R_{\rm RG}=199~R_\odot$ (thin black line).
Right ordinate indicates the ratio of the calculated X-ray flux 
to the original value due to partial/total eclipse by 
the RG and its tail. 
Vertical dashed line indicates the central time of X-ray 
total eclipses for both Model 1 and Model 2 ($\phi=0.075$). 
(b) Height of the RG and its tail from the orbital plane.   
Model 1 (thick red line). Model 2 (thin black line).
Red and black dotted lines denote the RG surface in Models 1
and 2, respectively.  Horizontal dashed line indicates
the radius of the X-ray-emitting sphere around the WD.
\label{X-ray}
}
\end{figure}

\section{X-RAY LIGHT CURVE} \label{sec_model}

\subsection{A Wide X-ray Eclipse}

Figure \ref{X-ray}a shows the X-ray fluxes of SMC3 against the orbital phase 
superposed for five orbital periods, the data for which are taken from
\citet{stu11}, 
assuming the ephemeris 
and period determined by Equation (\ref{equation_ephemeris}) 
in Section \ref{sec_Bmagmodel}.  
The same data are repeated for two orbital cycles.
This X-ray light curve shows the following characteristic properties. 

\begin{description}
\item[(1)] The X-ray eclipse is so wide that it spans
$\Delta \phi \sim$ 0.15--0.3  
of the orbital phase represented by Equation (\ref{equation_ephemeris}).
 
\item[(2)] The central time of X-ray eclipses is at $ \phi \sim 0.1$, i.e., 
$\sim 150$ days after $\phi=0.0$ as determined by Equation
(\ref{equation_ephemeris}), i.e., as determined from the long-term $V$ variation. 

\item[(3)] The eclipse has an asymmetric shape;  
the ingress is much steeper than the egress. 
\end{description}

Regarding property (1), the eclipse duration is not unusually long among 
symbiotic stars. The duration depends on the wavelength because of  
absorption/scattering by surrounding material (RG winds) outside 
the RG photosphere. For example, in the symbiotic nova PU Vul 
($P_{\rm orb}=13.5$ years), 
optical eclipse lasts 
$\Delta \phi=0.07$, which is longer than the RG geometrical eclipse 
$\Delta \phi=0.044$ because of TiO atmospheric absorption \citep{kat12}. 
At UV wavelengths region, its eclipse duration is even longer because of 
Rayleigh scattering as  $\Delta \phi > 0.14$ (up to 0.27)
\citep{tat09, kat11}.  

Regarding property (2), \citet{stu11} already reported that the X-ray minimum comes 
140 days after the $V$ magnitude minimum. They noted this as $B$ instead of $V$, 
but the MACHO blue is closer to yellow than $B$ \citep{bes99}. 
They attributed this difference between the minima to a 
possible error in the $V$ magnitude minimum 
due to the short observing period of MACHO (about 1.5 orbital periods). 
We see, however, a clear difference between the X-ray mid-eclipse 
and $V$ magnitude minima because we determine the ephemeris from
3.5 orbital periods as shown in
Figures \ref{BVlightcurve.08}--\ref{BVlightcurve.wd100}. 

All of these properties are consistent with an eclipse  
by an M giant companion swelling over the Roche lobe on the trailing
side of the binary orbit, as was already been suggested 
in the symbiotic star SY Mus \citep{dum99,per95}. 
The RG cool winds form a thick envelope, which tends to flow out
mainly backward (on the trailing side of the RG) owing to orbital motion.
As a result, SY Mus shows a wide eclipse in the UV continuum
($\phi \sim 0.9$ to $\sim 1.18$, i.e., $\Delta \phi =0.28$),
which is much wider than the geometric eclipse by the RG companion itself
($\Delta \phi = 0.074$).  Moreover, the UV 1325 \AA~continuum shows
an asymmetric eclipse in which the ingress is steeper and closer to the 
mid-eclipse than the egress. This property
is the same as property (3) of SMC3. 
\citet{dum99} analyzed the UV spectrum and concluded that these eclipse 
properties are naturally explained by obscuration of the UV flux by  
a thick neutral hydrogen cloud ($ N_{\rm H}$ from  
$10^{23}$~cm$^{-2}$ to $> 10^{25}$
~cm$^{-2}$) surrounding the RG.

Because the X-ray eclipse of SMC3 shows similar properties, 
we suppose that it is also caused   
not only by the RG companion itself, but also by a thick atmosphere
surrounding the RG.  In other words, properties (1)--(3) above are 
evidence of eclipses caused by a companion star swelling over the Roche
lobe on the trailing side of the binary orbit (see Figure \ref{binary.X}).

\subsection{Circumbinary Configuration} \label{sec_config}

For the X-ray emitting region, we simply assume 
a spherical hot nebula of radius $R_{\rm NE}$ centered on the WD.
The WD itself emits stronger X-rays than the surrounding nebula, 
but the WD is believed to be always obscured by a 
neutral hydrogen cloud surrounding the binary in the orbital plane. 
Therefore, we observe only the extended X-ray source (blue region 
in Figure \ref{binary.X}) which is always  
partially occulted by the trailing tail from the RG. 
We will discuss the X-ray emitting region 
in Section \ref{sec_discussion}. 

We assume that the surface brightness of the X-ray emitting region is 
constant. This may not be a good approximation, 
especially if the hot nebular gas 
originates in the wind from the WD, the density of which decreases 
as $r^{-2}$, where $r$ is the distance from the WD. 
\citet{ori07}, however, suggested that 
their spectral analysis with CLOUDY favors a constant density distribution 
over an $r^{-2}$ law. 
In that case, the surface emissivity (emissivity per unit
area of the hot nebula in the line of sight) is close 
to constant rather than decreasing with increasing power of the distance from 
the WD.  
We suppose that if the X-ray flux is due to Thomson 
scattering with an optical depth of about 1 
rather than thermal emission, the surface emissivity may not 
depend strongly on the distance from the WD. In this paper 
we do not discuss whether the 
X-ray flux is due to Thomson scattering, thermal emission 
of the hot plasma itself, or another cause, but, 
simply assume a uniform surface emissivity (a circular disk with 
a constant emissivity) as a first approximation.

Many numerical calculations show that cool RG winds (and also hot WD winds 
if present) form spiral structures around the binary 
\citep[][and references therein]{saw86,gaw02,kim07,edg08,moh12}. 
Ejected matter is highly aspherical, having a tendency to   
concentrate toward the orbital plane
\citep{bis00,gaw02,edg08,moh12}.
In optical spectra of SMC3 \citep{mor92,mue96,ori07},  
Raman-scattered \ion{O}{6} lines indicate 
a significant amount of neutral material \citep{shcm96}.
Therefore, we suppose that SMC3 has a tail of neutral material concentrated 
in the orbital plane.

Figure \ref{binary.X} shows a schematic picture of our circumbinary model 
of X-ray eclipses. The RG is swollen to almost twice in its size 
in the direction opposite to the orbital motion and is followed by a long, 
geometrically thick tail surrounding the binary in the orbital plane.  
The shape of the RG and its tail is drawn schematically using information 
from numerical calculations by \citet{saw86} and \citet{moh12}, which  
show narrow spiral tails flowing from both stars.
Mohamed \& Podsiadlowski's 3-D calculation showed 
that the outflowing matter is highly aspherical and concentrated in the orbital
plane.  The details of the mass distribution in the orbital plane 
do not affect our X-ray light curve model because we assume a total 
eclipse with an inclination of $i=90 ^\circ$. Therefore, only the heights 
(perpendicular to the orbital plane) of the RG and tail are 
important in our X-ray light curve modeling,  
which we use to reproduce the X-ray light curve. 
We suppose that this tail is composed of neutral hydrogen, which is  
opaque only to supersoft X-ray photons but transparent to optical 
photons. As shown later, the radius of the X-ray emitting region 
$R_{\rm NE}=100~R_\odot$, which is about half of the RG radius. 
This ratio is roughly consistent with the mass distribution around the 
WD in the 3-D calculation of binary mass outflow \citep{moh12}, 
although their binary parameter differs from ours.

Such neutral hydrogen material, opaque to X-rays but transparent to optical 
photons, was reported in the recurrent nova U Sco \citep{nes12}. 
In the supersoft X-ray phase followed by the 2010 outburst, U Sco temporarily 
showed irregular rectangular-shaped variations in the X-ray flux, whereas 
the UV and blue light curves showed a smooth eclipse. \citet{nes12} explained this X-ray 
variation as occultations of an extended X-ray region due to 
Thomson scattering by several blobs of neutral hydrogen moving in 
front of the X-ray emitting region. These blobs obscured supersoft X-rays 
but were transparent to UV/blue optical photons. 
In their analysis, the X-ray emitting region extends beyond 
the binary orbit (the orbital period is $P_{\rm orb}=1.23$ days, 
and the separation is $a=6.5 R_\odot$), and the blob sizes  
(0.73--1.1 $R_\odot$) are comparable to that of the 
companion MS star ($R_{\rm MS}=2.1~R_\odot$). Because 
SMC3 is a wide binary ($P_{\rm orb}=1634$ days, $a= 700$--1000 $R_\odot$ 
as tabulated in Table \ref{table_binary}) and its SSS phase 
lasts much longer than that of U Sco (which lasts only 15 days),   
the circumstellar matter could be much more massive and extend 
to a much larger size than in U Sco. 

In this paper, we assumed that the eclipse is essentially total; i.e., 
the X-ray emitting region (radius $R_{\rm NE}$) is completely 
covered by the RG, although weak X-ray fluxes 
($< 1/20$ of the X-ray high state) have been observed \citep{stu11}. 
We regard these weak X-rays as light scattered by tenuous plasma 
widely distributed beyond the radius $R_{\rm NE}~ ($=~100$~R_\odot)$.  
This assumption does not affect our light curve modeling because the 
scattered light is very faint.

\subsection{Model X-Ray Light Curve}

Our model, which reproduces the observed X-ray variation, 
is shown in Figure \ref{X-ray}. 
The orbital phase $\phi=0.0$ is defined as 
when the RG is in front of the WD.
The theoretical X-ray light curve is calculated from the area of 
the X-ray emitting region occulted by both the RG and its tail 
at a given orbital phase. 
Model 1 (denoted by a thick red line) assumes a greatly swollen RG with 
a thick (i.e., taller) tail and reproduces the lowest X-ray flux ever observed. 
On the other hand, Model 2 (denoted by a thin black line) assumes 
a smaller RG with a thin tail and reproduces the highest ever observed X-ray flux. 
The geometrical thickness of the tail is shown in 
Figure \ref{X-ray}b, and its minimum height is listed 
in Table \ref{table_eclipse}.

We found that an RG radius of 200--280 $R_\odot$ 
is consistent with the ingress steepness,  
the value of which depends weakly 
on $R_{\rm NE}$ and the height of the tail. 
The nebula size of $R_{\rm NE}=100~R_\odot$ may be a bit large, but 
it cannot be a point source compared with the RG  
because the ingress shows a finite slope that is not perpendicular.

Outside of total eclipses (i.e., at the orbital phase $\phi\sim0.3$--0.8), 
the X-ray data are scattered, as shown in Figure \ref{X-ray}a; 
this represents real variations between different orbital cycles   
or is attributed to possible errors in the conversion factor 
between different satellites \citep{stu11}. 
If this is a real variation, different X-ray fluxes are
attributed to different shapes (heights) of the RG tail.  
Because the RG is pulsating with a period of 110 days, the mass-loss rate 
of the cool wind could vary with time, in addition to exhibiting possible 
long-term irregular variations. Thus, the RG tail naturally has different 
thicknesses (heights) at different times.  
\citep[Note that periodic variations in  
the tail shape are observed in a 3-D calculation of the mass outflow
from Mira with a periodically changing mass-loss rate having
a period of 332 days:][]{moh12}. 
This explains the short-term variation in the X-ray flux 
as well as the differences among orbital periods.
Note that the combination of the RG size and geometrical 
thickness (height) of the tail is not unique. 
We determined the tail height relative to $R_{\rm NE}=100~R_\odot$. 
If we adopt a smaller $R_{\rm NE}$, for example, a  
similar X-ray light curve is obtained with the tail height 
multiplied by $R_{\rm NE}/(100~R_\odot)$ to the present models. 
The detailed distribution of circumbinary matter in the azimuthal 
and radial directions
on the orbital plane does not affect our X-ray eclipse light curve model 
because only the height of the tail at a given orbital phase is important
for determining the X-ray flux.

Figure \ref{X-ray}a also indicates the central time of the 
X-ray total eclipse by a dashed line, according to the ephemeris 
determined from the $V$ magnitude minimum. The X-ray mid-eclipse 
comes 123 days after $\phi=0.0$ determined from the minima in
the $V$ light curve.


\begin{figure}
\epsscale{1.15}
\plotone{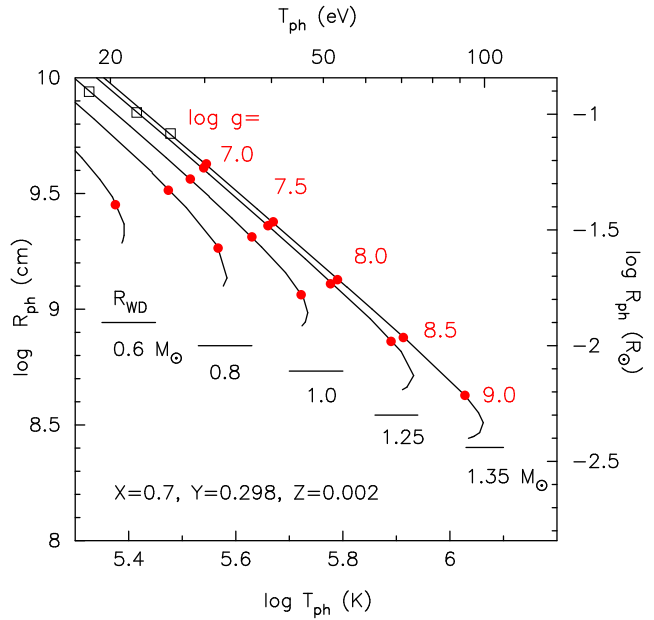}
\caption{Photospheric radius versus photospheric temperature 
of our envelope models for WDs with masses of 
0.6, 0.8, 1.0, 1.25, and 1.35 $M_\odot$. 
The chemical composition of the envelopes
is assumed to be $X=0.7$, $Y=0.298$, and $Z=0.002$. 
Red filled circles indicate places at $\log g $ = 7.0, 7.5, 
8.0, 8.5, and 9.0. 
Optically thick winds occur in the region left of open square. 
No optically thick winds are accelerated by the 0.6
and 0.8 $M_\odot$ WDs because of low metallicity in the SMC.
Short horizontal bars indicate the Chandrasekhar radii 
corresponding to each WD mass. 
\label{WDradius.T}
}
\end{figure}

\section{Discussion} \label{sec_discussion}

Figure \ref{WDradius.T} gives the photospheric radius versus the 
photospheric temperature for WDs of different masses  
with steady hydrogen-shell-burning.  
The chemical composition of the envelope is assumed to be $X=0.7,~Y=0.298$, 
and $Z=0.002$. The numerical method is the same as that
in \citet{kat99} and \citet{nom07}. 
In the region to the left of the open squares, optically thick winds occur  
and absorb supersoft X-rays. In the region to the right of 
the open squares, the WD envelope is in hydrostatic equilibrium, which 
corresponds to a SSS phase. 
The lowest end of each line (i.e., the smallest WD radius) corresponds to
the point at which hydrogen burning is extinguished.  \citet{ori07}
and \citet{stu11} obtained an X-ray temperature for SMC3 
as soft as 30--50 eV.  If this temperature corresponds directly to
the photospheric temperature, the radius of the photosphere is less 
than $0.1~R_\odot$, much smaller than the radius of the hot nebula,  
$R_{\rm NE}=100~R_\odot$. This is consistent with our assumption that 
the WD is always totally obscured by the RG and its long tail, and we
see only X-rays scattered by the hot nebula.

This figure also shows the positions of the WD envelopes
corresponding to surface gravities of $\log g$=7.0, 7.5, 8.0, 8.5,
and 9.0, where $g=GM_{\rm WD}/R_{\rm ph}^2$ is the photospheric gravity. 
There are no envelope solutions corresponding to $\log g$=8.5 and 9.0 
for $M_{\rm WD} \lesssim 1.0~M_\odot$. 
\citet{ori07} suggested a large gravity of $\log g \sim 9.0$ and a 
law temperature of 36--47 eV on the basis of X-ray spectral fitting. Such a large 
gravity corresponds, in our WD envelope models, to a very massive WD 
($\gtrsim 1.3~M_\odot$) with a small radius,
and also a much higher temperature 
($\gtrsim 100$ eV) as shown in Figure \ref{WDradius.T}.
The origin of this discrepancy is unclear.

On the basis of \citet{ori07}, we assumed that the WD is surrounded by a very hot nebula, 
which scatters X-rays from the WD. 
We estimate the density of the hot plasma from the coronal lines. 
In principle, the electron density in the coronal region should be close to the
critical density; e.g., the critical density for [\ion{Fe}{10}] 6374
is $4.8 \times 10^9$~cm$^{-3}$ \citep{nag01}.  In SMC3, both the  
[\ion{Fe}{10}] 6374 and [\ion{Fe}{9}] 7892 lines are relatively strong
and thus trace regions of high electron density,   
$N_{\rm e} \sim 10^9$--$10^{10}$ cm$^{-3}$. 
For a spherical emitting region of radius $R_{\rm NE}=100~R_\odot$ 
(fully ionized constant-density gas)
the total mass of the nebula is estimated to be roughly   
$10^{-9}$--$10^{-8}~ M_\odot$.  The optical depth of the region
in the line of sight is $\tau \sim \kappa \rho r =$ 0.004--0.04  
for the electron scattering opacity. Therefore, this region is optically thin, 
which is consistent with our assumption.  However, these values
are highly uncertain because of the simplified assumption.   

We also estimate the temperature of this region.
These coronal lines, [\ion{Fe}{10}] and [\ion{Fe}{9}], have high ionization 
potentials (IPs), of 234 eV and 262 eV, respectively, and their presence 
indicates that the ionized matter is at a high temperature. 
Using the formula $T/(10^3$ K) $\sim$ maximum observed IP (eV) 
\citep{mue94}, we obtain a temperature for the ionizing source of 
$T=$(2--3)$\times 10^5$ K. 
This plasma temperature is consistent with the WD temperature 
estimated from X-ray spectrum analysis, 36--47 eV 
(4.2--5.5 $\times 10^5$ K) \citep{ori07}.

\section{Conclusions} \label{sec_conclusions}

We estimated the radius of the RG and confirmed that it almost
fills its Roche lobe.  On the basis of this picture, we calculated 
the variations in orbital brightness including an irradiation effect and 
the geometrical effect of a tidally distorted companion. 
Our model reasonably reproduces the observed long-term $I$ and $V$ 
light curves of SMC3.  This again confirms that the RG almost
fills its Roche lobe.  Thus, the mass transfer rate from the RG to the
WD could be high enough ($\gtrsim$ a few times $10^{-7}~M_\sun$~yr$^{-1}$)
to maintain steady hydrogen-shell-burning on the WD \citep[e.g.,][]{nom07},
even though SMC3 is a long orbital period binary 
($P_{\rm orb}=4.5$ years).   
This is quite consistent with the fact that SMC3 has been a steady SSS 
for at least 20 years.

Our main results are summarized as follows. 
\begin{enumerate}
\item
We presented $I$ and $V$ light curve models of SMC3.
The observed magnitude 
variations can be reproduced by an irradiation of the RG 
by the hot WD as well as geometrical effects of the tidally distorted RG. 
This confirms that the RG component almost fills its Roche lobe, suggesting 
high mass transfer rates onto the WD, which maintain steady 
hydrogen-shell-burning on the WD. 

\item
We also presented an X-ray eclipse model  
in which an X-ray-emitting region is eclipsed by an RG companion 
swelling over the Roche lobe on the trailing side of the binary orbit. 
The X-ray-emitting region is highly ionized and extends to as large as
$\sim100~R_\sun$, scattering supersoft X-rays from the hot WD.
This configuration reasonably reproduces the asymmetry of the 
X-ray eclipses, that is, steeper ingress than egress and the occurrence of 
the center of the X-ray eclipse about 120 days after the $V$ minimum, 
as well as a very wide X-ray eclipse.  

\item
We also showed that our circumbinary matter configuration is
consistent with various observed characteristics; our configuration is composed of a  
very hot nebula surrounding the WD and circumbinary matter (a spiral tail)
of neutral hydrogen that always totally obscures the WD but partially
occults the extended X-ray emission region. These configurations may be
the key to the weakness of the supersoft X-ray flux in symbiotic systems. 
\end{enumerate}

\acknowledgments
We are grateful to the anonymous referee for useful  
comments that improved the manuscript. 
We also thank OGLE II/III for photometric data on SMC3. 
This paper utilizes public domain data obtained by the MACHO Project, jointly funded 
by the US Department of Energy through the University of California, 
Lawrence Livermore National Laboratory under contract No. W-7405-Eng-48, 
by the National Science Foundation through the Center for Particle Astrophysics 
of the University of California under cooperative agreement AST-8809616, 
and by the Mount Stromlo and Siding Spring Observatory, part of the Australian 
National University.
This research was supported in part by the
Grants-in-Aid for Scientific Research (22540254 and 24540227)
from the Japan Society for the Promotion of Science, 
and by the Polish National Science Center grant no. DEC-2011/01/B/ST9/06145.

\end{document}